\documentclass[aps,prl,twocolumn]{revtex4-2}
\usepackage{epsfig}
\usepackage{amsfonts}
\usepackage{amssymb}
\usepackage{enumitem}
\usepackage{amsmath}
\usepackage{amsthm}
\usepackage{subfig}
\usepackage{extarrows}
\usepackage{bm}
\usepackage{hyperref}
\usepackage{mathrsfs}
\usepackage{bbm}
\usepackage{cancel}
\usepackage[dvipsnames]{xcolor}
\usepackage{accents}
\usepackage{relsize}
\usepackage[capitalize]{cleveref}
  
  \crefname{section}{Sec.}{Secs.}
  \crefname{appendix}{App.}{Apps.}

\usepackage{xcolor}
\usepackage{hyperref}
\usepackage{mathtools}
\usepackage{amssymb,amsmath,graphics}   
\newcommand{\eq}{\begin{equation}}
\newcommand{\eqe}{\end{equation}}
\newcommand{\eqa}{\begin{eqnarray}}
\newcommand{\eqae}{\end{eqnarray}}

\begin{document}

\title{Emergence of Unitarity and Locality from Hidden Zeros at One-Loop Order}
\author{Jeffrey V. Backus$^1$}
\email{jvabackus@princeton.edu}
\author{Laurentiu Rodina$^2$}
\email{laurentiu.rodina@gmail.com}

\affiliation{$\mbox{}^{1}$Joseph Henry Laboratories, Princeton University, Princeton, NJ 08544, USA,}
\affiliation{$\mbox{}^{2}$Beijing Institute of Mathematical Sciences and Applications (BIMSA), Beijing, 101408, China}

\begin{abstract}
Recent investigations into the geometric structure of scattering amplitudes have revealed the surprising existence of ``hidden zeros'': secret kinematic loci where tree-level amplitudes in Tr$(\phi^3)$ theory, the Non-Linear Sigma Model (NLSM), and Yang-Mills theory vanish. In this Letter, we propose the extension of hidden zeros to one-loop-order in Tr$(\phi^3)$ theory and the NLSM using the ``surface integrand'' technology introduced by Arkani-Hamed et al. We demonstrate that, under the assumption of locality, one-loop integrands in Tr$(\phi^3)$ are unitary if and only if they satisfy these loop hidden zeros. We also present strong evidence that the hidden zeros themselves contain the constraints from locality, leading us to conjecture that the one-loop Tr$(\phi^3)$ integrand can be fixed by hidden zeros from a generically non-local, non-unitary ansatz. Near the one-loop zeros, we uncover a simple factorization behavior and conjecture that NLSM integrands are fixed by this property, also assuming neither locality nor unitarity. This work represents the first extension of such uniqueness results to loop integrands, demonstrating that locality and unitarity emerge from other principles even beyond leading order in perturbation theory.
\end{abstract}

 \maketitle
\section{Introduction}

Scattering amplitudes provide fascinating insights into the structure of quantum field theory (QFT). The bootstrap approach, which seeks to bypass traditional off-shell methods by determining amplitudes directly from imposing a set of basic physical principles, has emerged as a powerful framework in their study --- uncovering unexpected simplicity, efficient computational techniques, and novel mathematical structures. Such remarkable findings are persistent across a broad range of theories, from scalar theories like Tr$(\phi^3)$ and the Non-Linear Sigma Model (NLSM) to Yang-Mills and gravity. (See Refs.~\cite{Bern:2022jnl,Travaglini:2022uwo} for reviews.)
  
A natural objective within this bootstrap perspective is to identify the minimal set of constraints sufficient to uniquely determine scattering amplitudes. Recently, a series of uniqueness theorems  has revealed that the physical principles used to constrain amplitudes often constitute an \textit{overdetermined} system \cite{Arkani-Hamed:2016rak,Carrasco:2019qwr,Rodina:2016mbk,Rodina:2016jyz,Rodina:2018pcb,Rodina:2020jlw,Rodina:2024yfc}, implying that naively distinct physical principles are, in fact, \textit{not} independent. 
Most surprisingly, unitarity (and, in some cases, locality) were shown to emerge for free from other, purely on-shell principles.

The latest addition to this suite of defining principles is the ``hidden zeros'' \cite{Arkani-Hamed:2023swr}---specific kinematic configurations of external particle degrees of freedom where amplitudes vanish~\cite{osti_4736008,Zhou:2024ddy,Li:2024bwq,Huang:2025blb,Chang:2025cqe,Jones:2025rbv}. This property, totally obscured in the Feynman formulation, is remarkably generic (See, $e.g.$, Refs.~\cite{Kawai:1985xq,Bern:2008qj,Bern:2010ue,Bartsch:2024amu,Li:2024qfp,De:2025bmf}.). In earlier work \cite{Rodina:2024yfc}, it was proved that tree-level amplitudes in Tr$(\phi^3)$ theory are uniquely determined by hidden zeros under the assumption of locality, with unitarity an automatic consequence.

One of the greatest challenges in this uniqueness program has been extending these results beyond tree-level. If locality and unitarity are indeed to be regarded as emergent properties, it is crucial that we demonstrate this beyond leading order. However, finding such a uniqueness theorem at loop integrand level was previously considered not only difficult but fundamentally ill-defined. The notion of a unique loop integrand is ambiguous for two reasons: (1) internal loop momenta tailored to individual Feynman diagrams can be changed arbitrarily without affecting the physical amplitude, and (2) in massless theories, one manually tosses out ``1 / 0'' terms arising from tadpole and external bubble contributions, artificially destroying any single-loop cut structure in the integrand.

It is well-known that one can ameliorate the first issue by working in the planar limit. Recently, ``surface kinematics'' \cite{Arkani-Hamed:2023lbd,Arkani-Hamed:2024nhp,Arkani-Hamed:2024yvu,Arkani-Hamed:2024fyd,Arkani-Hamed:2024tzl,Arkani-Hamed:2024pzc,De:2024wsy,Salvatori:2018aha} has provided a solution to the second issue. With this canonical ``surface integrand,'' there is a natural means of accommodating contributions from tadpoles and external bubbles, in such a way that all single-loop cuts precisely match the expected tree amplitudes with totally generic kinematics.  

In this Letter, taking as example Tr($\phi^3$) theory and the NLSM, we show that these surface integrands can indeed be uniquely fixed at one-loop-order, with both locality and unitarity emerging from purely on-shell constraints. To do this, we first describe the kinematic mesh that neatly organizes kinematic data relevant for one-loop surface integrands \cite{Arkani-Hamed:2019vag,Arkani-Hamed:2019ttr}. We then propose the one-loop generalization of the hidden zeros discovered at tree-level in Ref.~\cite{Arkani-Hamed:2023swr}. Quite nicely, these zeros are maximal triangles (``big mountains'') on the one-loop mesh, in analogy to maximal rectangles on the tree-level mesh. In fact, our approach demonstrates that the various shapes of tree-level zeros have a \textit{unified} origin in different single-loop cuts of surface integrands. 

Near these big mountains, we describe a one-loop factorization pattern in both Tr$(\phi^3)$ and the NLSM, generalizing those found at tree-level in Ref.~\cite{Arkani-Hamed:2023swr}. This discovery adds to an ever-expanding list of factorization behaviors not anticipated from the requirements of unitarity \cite{Arkani-Hamed:2024fyd,Cachazo:2021wsz,Cao:2024qpp,Zhang:2024iun,Zhang:2024efe}. With these results in-hand, we prove that, under the assumption of locality, the Tr$(\phi^3)$ integrand is unitary \textit{if and only if} it satisfies the big mountain zeros. We then present strong evidence that big mountain zeros are still sufficient to uniquely determine the Tr$(\phi^3)$ integrand even without assuming locality. 

Finally, we find hidden zeros alone are not sufficient to fully fix the NLSM integrand, a conclusion similar to previous findings involving Adler-like zeros \cite{Bartsch:2022pyi,Bartsch:2024ofb}. Instead, we conjecture that, assuming neither locality nor unitarity, the stronger requirement of factorization near zeros uniquely fixes the NLSM integrand.

\section{Review: The kinematic mesh and zeros at tree-level}

In scalar theories, the kinematic data for an $n$-point scattering amplitude is specified by the momenta of all external particles $k_i^\mu$, satisfying momentum conservation and the on-shell condition. The scattering amplitude is then a rational function of Lorentz-invariant products of these momenta $k_i\cdot k_j$. In theories with \textit{colored} scalars, we can decompose the full amplitude into a sum over color-ordered amplitudes, each of which only receives contributions from planar diagrams. Any of these ``partial amplitudes'' can then be written solely in terms of the planar variables $X_{i,j}=(k_i+k_{i+1}+\ldots+k_{j-1})^2$. In the present work, we treat the $X$'s as a linearly-independent set; due to Gram determinant conditions, this implies that we are always working in $d > n -2$ dimensions for an $n$-point.

Non-planar invariants $c_{i,j}{=}{-} 2k_i{\cdot}k_j$ for $i, j$ not adjacent are related to the planar variables by the relation $c_{i,j}{=}X_{i,j}{+}X_{i+1,j+1}{-}X_{i,j+1}{-}X_{i+1,j}$.

\begin{figure}[t]
    \centering
    \includegraphics[width=0.8\linewidth]{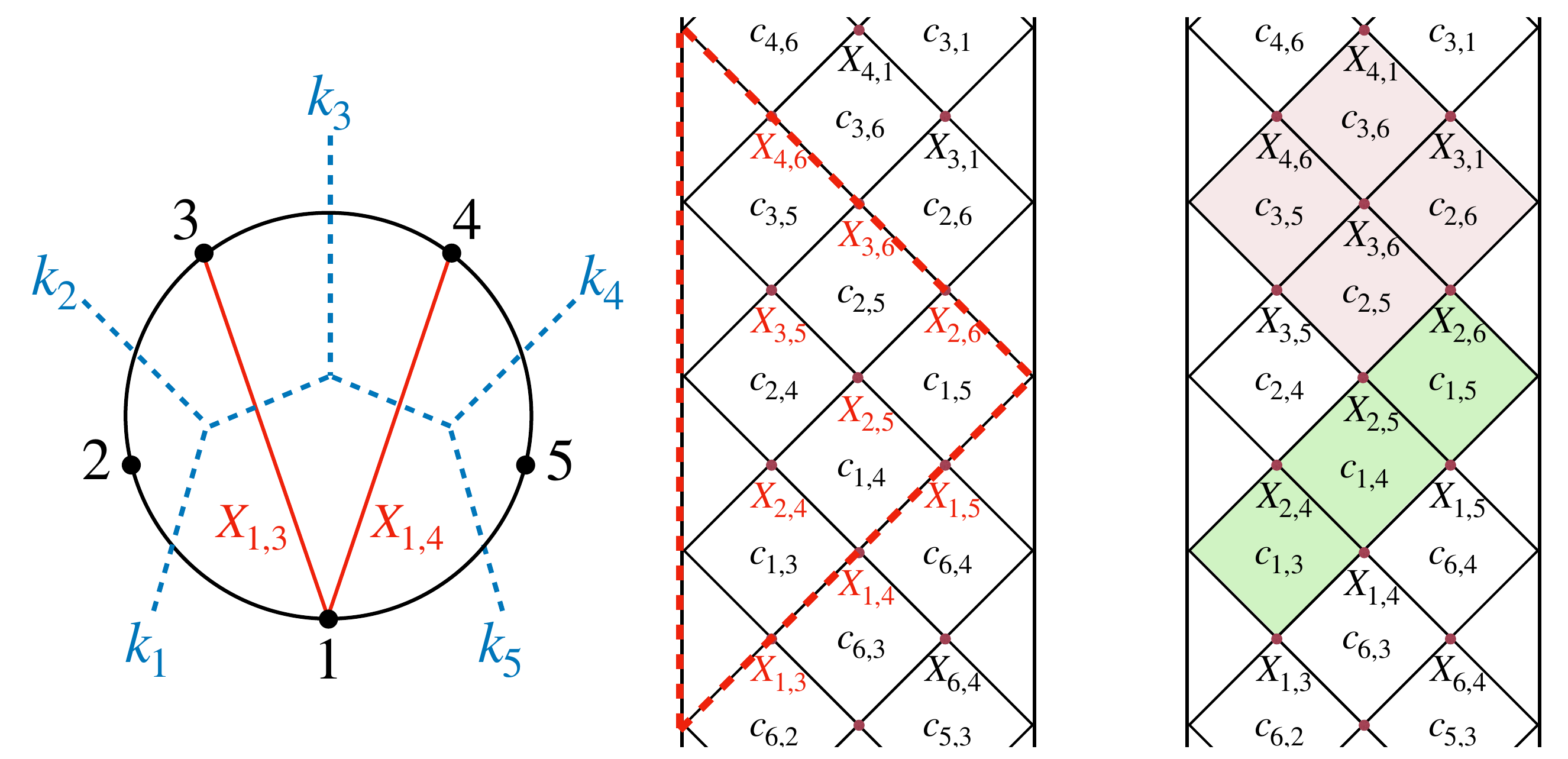}
    \caption{(From left to right) The momentum disk at five-points; the six-point kinematic mesh; examples of zeros on the mesh.}
    \label{fig:5point-pol+6pointmesh}
\end{figure}

The kinematic data for an \(n\)-point scattering process can be encoded in the
tree-level ``momentum disk'' (\textit{l.h.s.} of \cref{fig:5point-pol+6pointmesh}). 
The boundary has \(n\) marked points, with segments labeled by null external 
momenta. Each triangulation corresponds to a Tr\((\phi^3)\) Feynman diagram, 
with chords \(X_{i,j}\) as inverse propagators, and the amplitude follows by 
summing over all triangulations. NLSM amplitudes instead arise from 
even-angulations of the disk~\cite{Arkani-Hamed:2024yvu}.

Planar and non-planar data can also be organized using the tree-level 
``kinematic mesh''~\cite{Arkani-Hamed:2019vag} (\textit{r.h.s.} of 
\cref{fig:5point-pol+6pointmesh}). Each mesh vertex corresponds to an \(X\), 
and each diamond is labeled by a \(c_{i,j}\) whose indices match the \(X_{i,j}\) 
at its base. The relation for \(c_{i,j}\) is obtained by summing the two \(X\)’s 
at the top and bottom of the diamond and subtracting the two on the sides.

In Ref.~\cite{Arkani-Hamed:2023swr}, it was shown that color-ordered amplitudes in Tr\((\phi^3)\) and the NLSM vanish when all $c$'s in any maximal rectangle on the mesh are set to zero (the ``hidden zeros''). We can categorize these as $(k, m)$-zeros, each corresponding to the set:
\eqa
    \nonumber c_{i,j}&=&0,\ i\in\{m, m+1, \ldots, m + k - 1\}, \\  &j&\in\{k+m+1,k+m+2,\ldots, m - 2 + n\}\,.
\label{eq:one-zero}
\eqae
See \cref{fig:5point-pol+6pointmesh} for examples.

Locality requires all singularities to be simple poles in planar invariants $X_{i,j}$, with two $X$’s appearing together only if their chords on the momentum disk do not cross. Unitarity, via the optical theorem, demands that the residue on any pole factorizes into a product of lower-point amplitudes.

Ref.~\cite{Rodina:2024yfc} proved that imposing $n - 3$  $(1,m)$-zeros on a local ansatz restricts us to the Tr$(\phi^3)$ amplitude, up to an overall normalization. 
This ensures unitarity on every cut, showing that tree-level unitarity follows \textit{automatically} from hidden zeros and locality. 

\section{The Kinematic Mesh and Zeros at One-Loop}
\label{section:2}

Let us now see how this surface picture extends to loop-level. In scalar theories, the loop \textit{integrand} is a rational function of external and loop momenta. In the planar limit, one can construct a canonical one-loop \textit{surface} integrand in Tr$(\phi^3)$ and the NLSM endowed with both a momentum disk and a kinematic mesh, which we now describe.

\begin{figure}[t]
    \centering
    \includegraphics[width=1\linewidth]{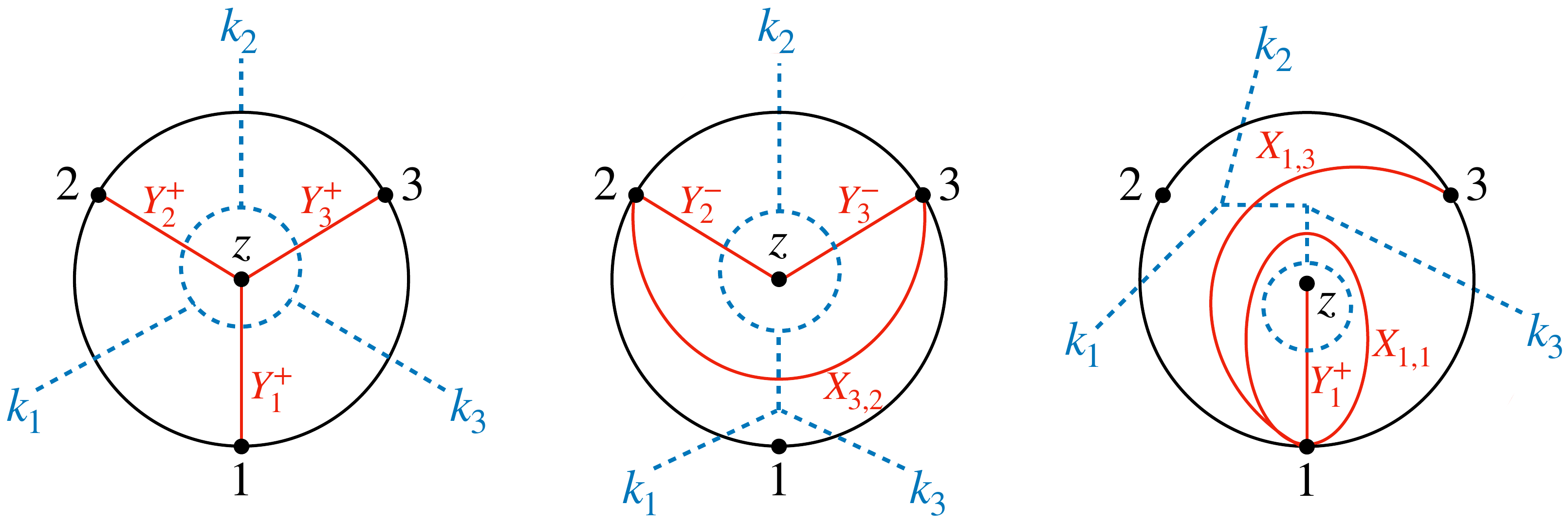}
    \caption{The one-loop surface picture for kinematic variables and dual Feynman diagrams in Tr$(\phi^3)$ theory, including tadpoles (right) and external bubbles (middle).}
    \label{fig:polygon}
\end{figure}

At one-loop, the analog of the tree-level momentum disk is the \textit{punctured} momentum disk, as shown in \cref{fig:polygon}. The inclusion of the puncture $z$ introduces some new properties of chords on this surface not present at tree-level but critical to describing the surface integrand. For example, using the convention that $X_{i,j}$ denotes a chord that wraps around the puncture in increasing order from $i$ to $j$ (when $i,j \neq z$), it is clear from the surface picture that $X_{i,j} \neq X_{j,i}$, even though naively by ``momentum homology'' these two should be equal. (We say that two curves are identified via momentum homology when, upon deforming both curves to the boundary of the disk, they are assigned the same momentum dependence. Hence, $X_{i,j}$ and $X_{j,i}$ are momentum homologous, precisely because the puncture does not carry any momentum.) What's more, the surface includes curves that obviously \textit{vanish} by homology, such as the chords $X_{i,i}$ that start on $i$, wrap around the puncture once, and return to $i$. (See the \textit{r.h.s.} of \cref{fig:polygon} for an example.) 

On the other hand, the puncture introduces new variables $X_{i,z} = Y_i$ whose curves start on $i$ and end on the puncture. These are the variables that, when using homology, depend on the internal loop momentum $l^\mu$: fixing, $e.g.$, $Y_1 = l^2$, we have $Y_i = (l + p_1 + \ldots + p_{i-1})^2$ for the rest. For reasons that will become clear in a moment, we will also need to equip the puncture with a parity $Y_i \to Y_i^\pm$, and then take the sum of the integrand with all loop variables positive $Y_i^+$ and the integrand with all loop variables negative $Y_i^-$ \cite{Arkani-Hamed:2019vag,Arkani-Hamed:2019ttr}. 

The surface integrand is defined as a function of this extended kinematic space \cite{Arkani-Hamed:2024tzl}. This means that we will need to work with variables that vanish in momentum space: each triangulation with a curve $X_{i,i}$ has a tadpole, and those with $X_{i+1,i}$ (and no $X_{i,i}$) have massless external bubbles. To get back to the familiar integrand in momentum space, we can always take away the parity of the puncture, throw out scaleless contributions by hand, set $X_{i,j} = X_{j,i}$, and express the $X_{i,j}$ and ${Y_i}$ in terms of external momenta $k_i^\mu$ and the loop momentum $l^\mu$ via homology. See Appx. F for examples of surface integrands.

\begin{figure}[t]
    \centering
    \includegraphics[width=0.8\linewidth]{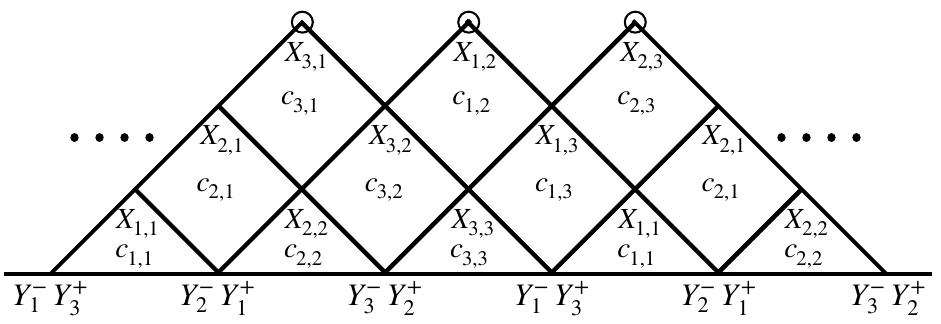}
    \caption{The $Y^- Y^+$ mesh at three points.}
    \label{fig:three-pt-one-loop-mesh}
\end{figure}

One can again organize the surface kinematic data using a mesh, an example of which is shown in \cref{fig:three-pt-one-loop-mesh}. Note that the propagator doubling $Y_i \to Y_i^\pm$ is  crucial to the construction of the mesh. As at tree-level, each (half) diamond in the one-loop mesh corresponds to a non-planar invariant $c_{i,j}$ whose indices now match those of the planar variable $X_{i,j}$ at the \textit{top} of the diamond. The relationship between the $c$'s and the $X$'s is the same as at tree-level, except for $c_{i,i}$ and $c_{i+1,i}$, where we find
\eqa
\label{eq:c-X-loop-1}
    \nonumber c_{i,i}&=&X_{i,i}-{Y}^-_i-{Y}^+_i,\\
    c_{i+1,i}&=&X_{i+1,i}-X_{i,i}-X_{i+1,i+1}+{Y}^-_{i+1}+{Y}^+_i\, .
\eqae

The zeros on this mesh correspond to setting all $c$'s in a maximal triangle (a ``big mountain'') to zero. That is, to impose the zero corresponding to the peak labeled by $c_{i-1,i} = 0$, we need to set
\eq
    c_{m,k}=0,\ m \in \{k, k+1,\ldots, i - 1\},
\label{eq:big-mountain-zero}
\eqe
for each $k = 1, 2, \ldots, n$. Since these zeros can only be phrased with surface kinematics, their existence is totally inaccessible in a generic integrand written in terms of momenta. In Appx. E, we give the proof that these configurations force integrands in Tr$(\phi^3)$ and the NLSM to vanish.

Note that we can equally define a mesh with ``$-$'' and ``$+$'' switched in \cref{fig:three-pt-one-loop-mesh}, leading to another set of hidden zeros. Hence, the $n$-point one-loop integrand has $2n$ big mountain zeros. We will denote the zero with peak at $c_{i-1,i} = 0$ and ordering $Y^\pm Y^\mp$ as the $(i, \pm)$-zero of the one-loop mesh. But, there is really a \textit{degeneracy} between the two parities: the integrand satisfies all $(i, +)$-zeros if and only if it satisfies all $(i, -)$-zeros. (See Appx. E for the proof.)

As emphasized above, one of the central novelties that surface kinematics buys us is well-defined single-loop cuts. On the punctured disk (as in \cref{fig:res-kin-disc}), a single-loop cut turns chords of an $n$-point one-loop integrand into chords of an $(n+2)$-point tree amplitude. Complementarily, on the one-loop kinematic mesh, this type of cut isolates a triangular region, corresponding precisely to a ray triangulation of the tree-level $(n+2)$-point mesh. (See \cref{fig:res-kin-mesh}.) As we will demonstrate later, the intersection of this triangle with a one-loop mountain zero yields a maximal rectangle, identifying a tree-level zero; see \cref{fig:residue}.

Finally, in the surface integrand picture, locality forbids crossing chords on the punctured disk, and unitarity demands consistent factorization on cuts through any set of chords \footnote{While these definitions guarantee that the physical loop-integrated amplitude is consistent with locality and unitarity, they constitute strictly stronger requirements, as they include rules for dealing with tadpoles and external bubbles that do not appear post-loop-integration.}.

\subsection{Factorization Near Zeros}

In Ref.~\cite{Arkani-Hamed:2023swr}, it was shown that, when setting all $c_{i,j}$ to zero in a particular $(k,m)$-zero of the tree-level mesh except one $c_{\star} \neq 0$, the amplitude in either Tr$(\phi^3)$ or the NLSM factorizes into two sub-amplitudes times an overall simple prefactor.

We uncover a similar story at one-loop. The mesh itself suggests a natural factorization: deleting any large mountain zero leaves a ``ray triangulation'' slice of the \((n+2)\)-point tree mesh, with the mountain's sides forming its boundaries. Since turning on any \(c_\star \neq 0\) inside an \((i,\mp)\)-zero enforces  $c_\star = -\bigl(Y_i^\mp + Y_{i-1}^\pm\bigr)$, this points to a factorization of the Tr\((\phi^3)\) and NLSM integrands as
\eq
\label{eq:fact-near-zero-int}
\mathcal{I}_n\left(c_{\star} \neq 0\right)=\left( \frac{1}{Y_i^\mp} + \frac{1}{Y_{i-1}^\pm} \right) \times \mathcal{A}_{n+2}\,,
\eqe
with tree-level amplitude $\mathcal{A}_{n+2}$ depending on the kinematics in the remaining tree-level mesh with some modifications on its upper and lower boundaries. 

\begin{figure}[t]
    \centering
    \includegraphics[width=0.9\linewidth]{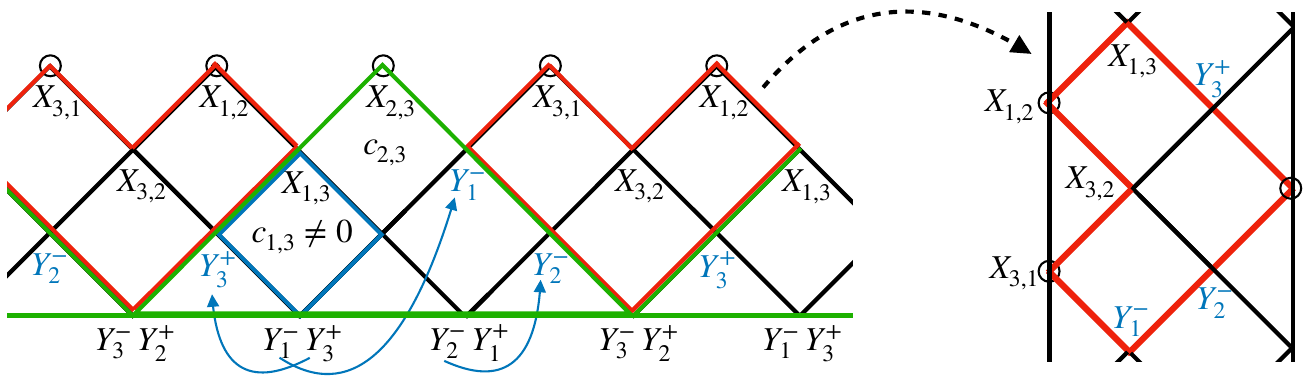}
    \caption{Factorization of variables near the $(3,-)$-zero at three-points. The higher point tree amplitude corresponds to the red-highlighted triangular region. Its kinematic dependence requires replacing some $X$ variables with $Y$ variables, indicated by the blue arrows.}
    \label{fig:factor-example}
\end{figure}

For Tr$(\phi^3)$, this holds exactly. For the NLSM, the picture is subtler. If we treat minus as odd and plus as even in the $\delta$-shift (see Ref.~\cite{Arkani-Hamed:2024nhp} for details on the $\delta$-shift, and Ref.~\cite{Paranjape:2025wjk} for a recent extension), the integrand factorizes as in \cref{eq:fact-near-zero-int} near $(i,-)$-zeros for even $i$ and near $(i,+)$-zeros for odd $i$. Factorization at the other zeros requires the opposite-parity prescription (minus even, plus odd). See Fig.~\ref{fig:factor-example} for an example, and Fig.~\ref{fig:factor} for the general case. We prove this pattern in Appx.~A.

\section{Uniqueness from zeros at One-Loop in Tr($\phi^3$)}

\subsection{Assuming Locality}

Let us now prove that the one-loop mesh zeros uniquely fix the color-ordered, \textit{local} (but non-unitary) ansatz for the surface integrand $\mathcal{M}_n$ in Tr$(\phi^3)$ theory. Below we present only the main steps.

We will start with the same generically non-unitary ansatz as in Ref.~\cite{Rodina:2024yfc}: an arbitrary linear combination of all full triangulations of the punctured disk. Taking a single-loop cut of $\mathcal{M}_n$ gives us a non-unitary ansatz for an $(n + 2)$-point tree-level amplitude $\mathcal{B}_{n+2}$:
\eq
\textrm{Res}_{Y=0}\mathcal{M}_n=\mathcal{B}_{n+2}\,.
\eqe

In Appx. B, we prove that imposing a big mountain zero on $\mathcal{M}_n$ \textit{secretly} imposes a tree-level zero on $\mathcal{B}_{n+2}$. Concretely, we show that, for the cut on $Y_j^+$, an $(i,-)$-loop-zero acts as an $(i - j - 1,j)$-tree-zero for any $i = j + 2, j + 3, \ldots, n + j$. That is, the different zero types at tree-level all originate from various combinations of loop zeros and single-loop cuts! Graphically, this can be observed very nicely on the kinematic mesh: the overlap between a mountain zero and the tree mesh induced by the cut is precisely a maximal rectangle, corresponding to a tree-level zero, see \cref{fig:residue}.

Then, we find that this set of zeros imposes identical constraints on $\mathcal{B}_{n+2}$ as $(n + 2) - 3$ distinct $(1,m)$-zeros, which were revealed in Ref.~\cite{Rodina:2024yfc} to be sufficient to fix $\mathcal{B}_{n+2}$ to be the \textit{unitary} amplitude $\mathcal{A}_{n+2}$ up to an overall number. See Appx. C for the proof of the equivalence between the two sets of zeros.

Next, overlap among all cuts with the same parity $Y^+$ arises from the $1/{Y_1^+\cdots Y_n^+}$ term in $\mathcal{M}_n$, forcing all positive-parity terms to carry the same weight $a^+$. By identical reasoning, negative-parity terms share weight $a^-$. Since every term in $\mathcal{M}_n$ contains at least one $Y$, the ansatz reduces to
\begin{equation}
\mathcal{M}_{n} \;=\; a^+ \,\mathcal{I}_{n}^+ \;+\; a^- \,\mathcal{I}_{n}^- \,,
\end{equation}
where $\mathcal{I}_n^\pm$ denotes the $n$-point integrand with only $Y^\pm$ poles. In Appx. D, we prove that $a^+=a^-$ by imposing the \((2,-)\)-zero and cuting \( Y_1^+ \), a configuration where the tree mesh avoids overlap with the mountain zero. 

This proves that $n$ mountain zeros uniquely fix the local ansatz for the one-loop integrand in Tr$(\phi^3)$ theory; therefore, unitarity follows as an automatic consequence of our assumptions. Since the number of weights in our one-loop ansatz grows exponentially with $n$, the emergence of unitarity from $n$ hidden zeros and locality is remarkably non-trivial. It turns out that this is really an \textit{equivalence}; see Appx. E, where we prove that unitarity implies the big mountain zeros.

\subsection{Without Assuming Locality}

Dropping locality, we take an ansatz of all possible $n$-pole terms built from $X_{i,j}$ or $Y_i^\pm$, no longer corresponding to mesh triangulations, but forbidding factors of both $Y^-$ and $Y^+$ in the same term. This ansatz grows \textit{factorially} with multiplicity, containing $\sim 6,500$ terms at four-points. We have verified that big mountain zeros fully fix this one-loop integrand up to four-points, allowing us to conjecture that locality and unitarity \textit{both} emerge from hidden zeros at loop level!

\section{Uniqueness from Factorization Near Zeros in the NLSM}

We now study uniqueness of the NLSM one-loop integrand. The non-local, non-unitary $2n$-point ansatz includes all terms built from any $n$ $X$ or $Y$ variables in numerator and denominator, forbidding opposite-parity $Y$’s in the same term. At four-points, the 136 possible $XX$, $XY$, and $YY$ factors generate $\sim 25{,}000$ terms.

From four-points onward, zeros alone fail to fix the integrand \footnote{For the purposes of constructing amplitudes, one practical solution to obtain a unique result is to allow only same-parity terms in the numerator, $i.e.$, only $X_{e,e}$, $X_{o,o}$, and $Y^{+}_e$ and $Y^{-}_o$ (when, for example, assigning $Y^{+/-}$ to be even/odd parity). This eliminates the trivial solutions and gives an efficient bootstrap procedure.}. Instead, we test factorization near zeros. For all $(i, \mp)$-zeros (where plus corresponds to $i$ odd and minus to $i$ even), we require the ansatz satisfies a factorization of the type
\eq
\label{eq.fix_fact}
\mathcal{M}_n(c_*\neq 0)\rightarrow \left(\frac{1}{Y_i^\mp} + \frac{1}{Y_{i-1}^\pm}\right)\times \mathcal{B}^{i,c_\star}_{n+2}(\tilde{X})\,,
\eqe
for all $c_*\neq 0$ within those zeros. Here, $B^{i,c_*}_{n+2}(\tilde{X})$ are general ansatze for the $(n+2)$-point tree NLSM amplitude, different for each factorization, and the set $\tilde{X}$ is the $X,Y$ variables consistent with the graphical rule in \cref{fig:factor}. 

At four-point, we have verified that these constraints uniquely reproduce the NLSM integrand, also fixing the \textit{l.h.s.} amplitudes in \cref{eq.fix_fact} in the process. This strongly suggests that factorization near zeros fully determines NLSM integrands at all multiplicities, with both locality and unitarity emerging from this property.

\section{Outlook}
In this Letter, we found that one-loop integrands in two scalar theories are uniquely determined by imposing novel constraints: loop hidden zeros and factorization near these zeros. Since our approach assumes neither locality nor unitarity, we have shown that these two seemingly fundamental principles of QFT can be viewed as emergent, even beyond tree-level.

Strictly speaking, our proof applies only to a local Tr\((\phi^3)\) ansatz, where we uncovered the unification of all tree-level zeros in single-loop cuts of the integrand. A natural next step is to prove the same holds even without assuming locality. Another important question is whether any structure of the loop-level zeros survives when returning from surface kinematics to momentum space. Identifying such a feature would likely be nontrivial, but of significant interest.

The surface kinematics framework applies at any loop order \cite{Arkani-Hamed:2024pzc}, and it can also be used to write down a Yang-Mills integrand with well-defined  single-loop cuts \cite{Arkani-Hamed:2024tzl,Cao:2025mlt,Backus:2025njt}. Thus, a broad range of tests involving uniqueness and emergence from hidden zeros and factorization is now a tangible next step--- beyond scalar theories, and at all orders in perturbation theory.

\vspace{\baselineskip}

\begin{acknowledgments} 
{\it Acknowledgments}---It is our pleasure to thank Carolina Figueiredo and Giulio Salvatori for insightful discussions and Nima Arkani-Hamed for helpful comments on the draft. JB is supported by the NSF Graduate Research Fellowship under Grant No. KB0013612. LR is supported by the Beijing Natural Science Foundation International Scientist Project No. IS24014, and the National Natural Science Foundation of China General Program No. 12475070.
\end{acknowledgments}

\appendix
\section{Appendix A: Factorization proof}\label{appx.A}

Here, we prove the factorization formula~\eqref{eq:fact-near-zero-int}. To do this, let us fix a particular $(i, -)$-zero and $c_{a,b} \neq 0$ (the ``loop near-zero'' condition) and impose them on the Tr$(\phi^3)$ integrand $\mathcal{I}_n$. We can then cut on \textit{any} $Y_j^\pm$, which will produce a particular tree amplitude $\mathcal{A}'_{n+2}$ with generic kinematics on the support of the loop near-zero. For the moment, let us restrict so that we cut on neither $Y_i^-$ nor $Y_{i-1}^+$. In this case, after imposing the loop near-zero condition, there are two options for $\mathcal{A}'_{n+2}$. If the overlap between the loop and tree meshes (as shown in \cref{fig:residue}) contains only $c$ variables that vanish, then $\mathcal{A}'_{n+2} = 0$. 

If, alternatively, the overlap contains $c_{a,b} \neq 0$, the tree amplitude will factorize as 
\eq
    \mathcal{A}'_{n+2}=\left(\frac{c_{a,b}}{X_{\mathrm{B}}X_{\mathrm{T}}}\right) \times \mathcal{A}_{\text {down }} \times \mathcal{A}_{\mathrm{up}}\,,
\eqe
with $r.h.s.$ kinematics given by the rules in Ref.~\cite{Arkani-Hamed:2023swr}. Looking at \cref{fig:residue}, we can easily pick out the universal prefactor for the factorization of $\mathcal{A}'_{n+2}$ in this loop near-zero configuration: it is the four-point tree amplitude with kinematics given by the left-most and right-most corners of the overlap region. When cutting on a $Y_j^+$, the right-most planar variable is always $Y_{i - 1}^+$. The left-most is $X_{j,i} = Y_i^- + Y_j^+ = Y_i^-$ on the cut. (The same story, with sides reversed, holds for cuts on $Y_j^-$.) So, we have
\eq
    \mathcal{A}'_{n+2} \to \left( \frac{1}{Y_i^-} + \frac{1}{Y_{i -1}^+} \right) \times \mathcal{R}\,,
\eqe
where $\mathcal{R}$ is the product of unimportant lower-point amplitudes. This implies that all terms in $\mathcal{I}_n$ with at least one $Y$ that is neither $Y_i^-$ nor $Y_{i-1}^+$ is proportional to the prefactor shown in \cref{eq:fact-near-zero-int}. 

\begin{figure}[t]
    \centering
    \includegraphics[width=00.9\linewidth]{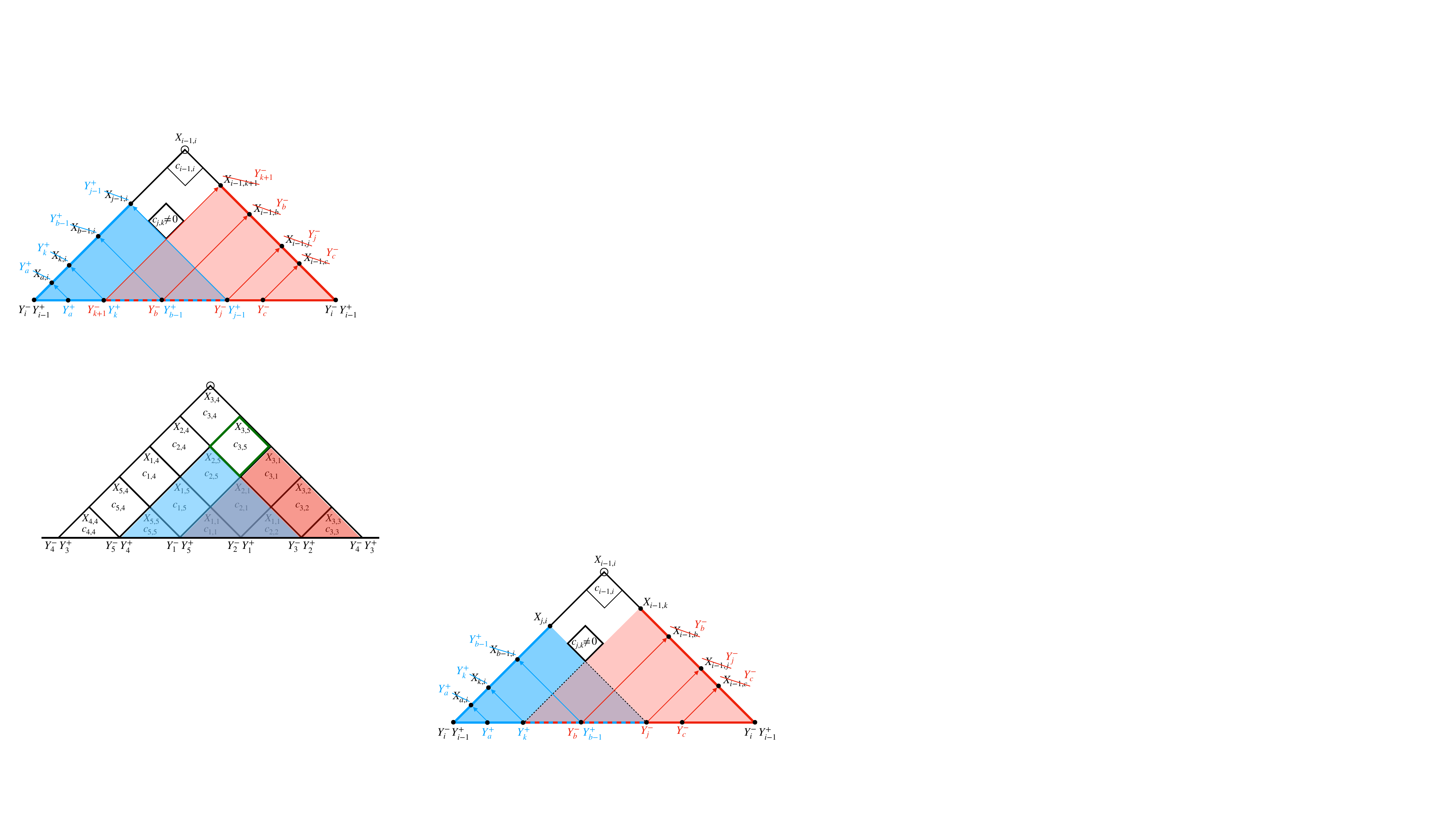}
    \caption{Factorization of variables near a $(i , -)$-zero with $c_{j,k} \neq 0$. The region below and beside $c_{j,k}$ is divided into blue and red parts. All the $X$ variables on the slope of the mountain in a colored region are replaced with $Y$ variables using the rule shown in the figure.}
    \label{fig:factor}
\end{figure}

When we now cut on $Y_{i}^-$, we are left with a tree amplitude $\mathcal{A}^{(1)}_{n+2}$ whose mesh has no overlap with the zero. As such, it will neither vanish nor factorize. As we discuss in Appx. B, the kinematics of $\mathcal{A}^{(1)}_{n+2}$ are precisely those in the tree mesh formed as in \cref{fig:res-kin-mesh} but instead with a shift on its left edge $X_{i-1,m} \to Y_m^-$ for all $m$. However, on the loop near-zero and cut, we also have $X_{m,i} = Y_m^+$ for all $m \leq a - 1$ and $Y_m^- = X_{i-1,m}$ when $m < b+1$. Note that this derives, respectively, the rules in the blue and red triangles in \cref{fig:factor}. We can proceed with the same arguments for the cut on $Y_{i-1}^+$, obtaining $\mathcal{A}^{(2)}_{n+2}$ whose tree mesh has the same interior as that of $\mathcal{A}^{(1)}_{n+2}$ and also undergoes the shifts drawn in \cref{fig:factor} on the edges. So, $\mathcal{A}_{n+2} = \mathcal{A}^{(1)}_{n+2} = \mathcal{A}^{(2)}_{n+2}$ on this kinematic configuration.

Thus, we have shown that, on a loop near-zero, the integrand is proportional to the prefactor given \cref{eq:fact-near-zero-int}, and that this prefactor multiples the tree amplitude $\mathcal{A}_{n+2}$ with kinematics as shown in \cref{fig:factor}. So, we have proved \cref{eq:fact-near-zero-int} in Tr$(\phi^3)$ theory. As always, the proof for $(i, +)$-zeros goes through in an identical manner.

In Ref.~\cite{Arkani-Hamed:2023swr}, the analogous tree-level factorization was proved using classic ``stringy'' integrals that UV-complete Tr$(\phi^3)$ and NLSM tree-level amplitudes. In this work, we instead prove this factorization solely from from properties of the ``low-energy'' field-theory object $\mathcal{I}_n$.

\section{Appendix B: Loop-zeros impose tree-zeros on the single-loop cuts}

In this Appendix, we prove that the loop-zeros act as tree-zeros on single-loop cuts of the loop-integrand ansatz. We'll consider here the $(i, -)$-zeros and the $Y^+$ cuts. (Different parity choices can be treated identically.) For simplicity, we'll focus on cutting $Y_1^+$. The first thing we must do is prove that the operations of imposing a zero and taking a residue commute, since our assumption is the full integrand vanishes (not individual cuts). As long as $i \neq 2$, we find that
\eq
\label{eq:ops-comm}
\textrm{Res}_{Y^+_1=0}(\mathcal{M}_{n}|_{(i,-)})=\left(\textrm{Res}_{Y^+_1=0}\mathcal{M}_{n}\right)_{(i,-),Y^+_1=0}\,.
\eqe
This is because, for $i \neq 2$, setting $Y_1^+ = 0$ on the support of an $(i, -)$-zero does not force any other kinematic variables to vanish, ensuring the residues on both sides of \cref{eq:ops-comm} pick up the same terms. To contrast, for $i=2$ the zero sets $Y_{2}^-=-Y_1^+$, so the residue on the \textit{l.h.s.} will pick up extra terms as compared to the \textit{r.h.s.} corresponding to the pole in $Y_{2}^-$.

\begin{figure}[h]
    \centering
\includegraphics[width=1\linewidth]{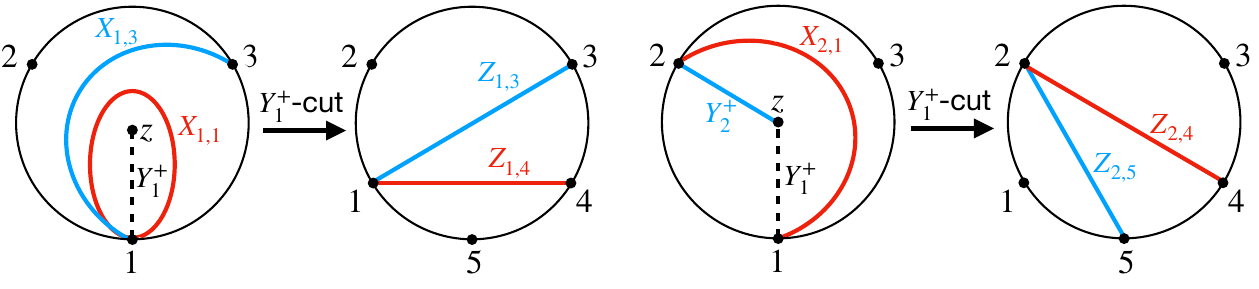}
    \caption{The $Y_1^{+}$ cut of an $n$-point integrand induces a local triangulation of the $(n+2)$-point tree-disk.}
    \label{fig:res-kin-disc}
\end{figure}

Because it is our hypothesis that $\mathcal{M}_n$ satisfies the big mountain zeros, we have the condition
\eq
\label{eq:loop-cut-van}
\left( \textrm{Res}_{Y^+_1=0}(\mathcal{M}_{n})\right)_{(i,-),Y^+_1=0} = \mathcal{B}_{n+2}(Z_{i,j})|_{(i,-),Y^+_1=0} = 0\,,
\eqe
for all $i \neq 2$, where $\mathcal{B}_{n+2}$ is our local tree-level ansatz with generic kinematics we denote by $Z_{i,j}$ to distinguish them from the loop $X_{i,j}$. It is simple to work out the mapping between the $Z$'s and the $X$/$Y$'s: the non-trivial ones are $Y_i^+\leftrightarrow Z_{i,n+2} = Z_{n+2,i} $, $X_{i,1}\leftrightarrow Z_{i,n+1} = Z_{n+1,i} $, and  $X_{1,i}\leftrightarrow Z_{i,1} = Z_{1,i} $ for $i = 2, 3, \ldots, n$, and $X_{1,1}\leftrightarrow Z_{1,n+1}  $. For the rest, we have $X_{i,j}\leftrightarrow Z_{i,j} = Z_{j,i} $ for $i < j$. (Note that $X_{i,j}$ with $i \ge j$ are always locally inconsistent with the $Y_1^+$ chord.) We show examples of how this works in \cref{fig:res-kin-disc}.

\begin{figure}[h]
    \centering
\includegraphics[width=0.55\linewidth]{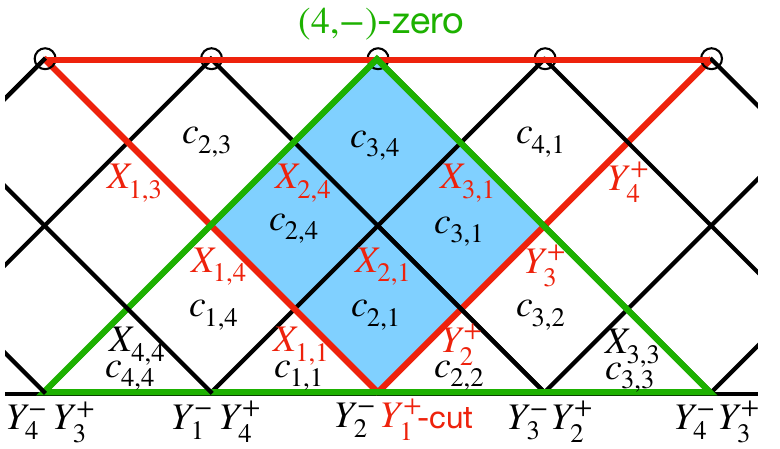}
    \caption{The mountain zeros (green) intersect the mesh area corresponding to a tree amplitude (red) in maximal rectangular regions, producing the tree-level zeros.}
    \label{fig:residue}
\end{figure}

\begin{figure}[h]
    \centering
\includegraphics[width=0.8\linewidth]{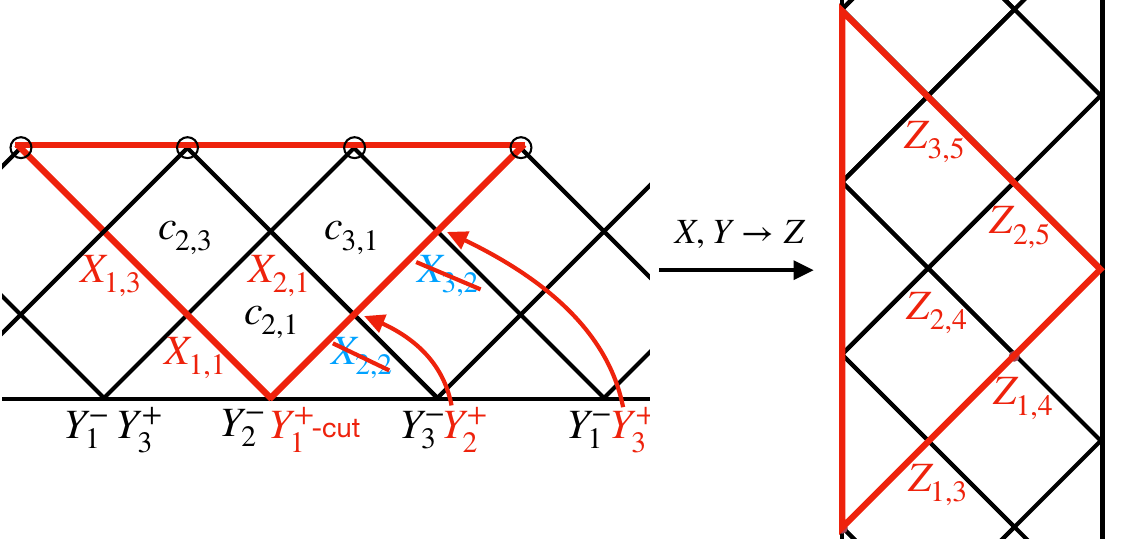}
    \caption{The $Y_1^{+}$ cut defines the ray triangulation of an $(n+2)$-point tree mesh, where the loop-level $X_{i,2}$ on the right edge are replaced with the corresponding $Y^+_i$.}
    \label{fig:res-kin-mesh}
\end{figure}

Now, the local structure of $\mathcal{B}_{n+2}$ in terms of $Z$ variables defines a tree-level mesh which can easily be obtained graphically from the loop mesh, as we show in \cref{fig:res-kin-mesh}. Quite beautifully, for any $i \neq 2$, the overlap of an $(i, -)$-zero and the tree mesh of $\mathcal{B}_{n+2}$ is exactly a maximal rectangle--- a tree-level hidden zero! (See the blue diamond in \cref{fig:residue}.)

For diamonds that do not touch the right edge of the tree mesh, it is clear that the $c_{i,j}(X)=0$ conditions of the loop zero trivially induce the corresponding $c_{i,j}(Z)=0$ conditions in the tree mesh. For diamonds that do touch the right edge (but are not the bottom-most diamond), the mountain zero condition $c_{k,1}(X) = 0$ implies that
\eq
\label{eq:lin-rel}
    X_{k,1} + X_{k-1,2} = X_{k-1,1} + X_{k,2}\,.
\eqe
But, we also have $X_{i,2} = Y_i^+ + Y_2^-$ for all $i \leq \textrm{max}(k)$ on the support of the zero. So, \cref{eq:lin-rel} is equivalent to $X_{k,1} + Y_{k-1}^+ = X_{k-1,1} + Y_k^+$, which is exactly what appears on the overlap between the tree-mesh and loop-mesh in \cref{fig:residue}. Finally, for the bottom-most diamond, setting $c_{2,1} = 0$ and $Y_1^+ = 0$ tells us
\eq
\label{eq:lin-rel-2}
    X_{2,1} + Y_2^- = X_{1,1} + X_{2,2}\,.
\eqe
Since we have $X_{2,2} = Y_2^+ + Y_2^-$ on the support of the zero, this condition matches what is shown in \cref{fig:residue}.

As a result of these arguments, the fact that $\mathcal{B}_{n+2}$ vanishes on the zero+cut condition shown in \cref{eq:loop-cut-van} is \textit{equivalent} to it vanishing on a tree-level hidden zero! In general, for the cut on $Y_j^+$, an $(i,-)$-loop-zero acts as an $(i - j - 1,j)$-tree-zero for all $i = j + 2, j + 3, \ldots, n + j$.

\section{Appendix C: Tree-level uniqueness from diverse zeros}\label{appx.C}
In this Appendix, we show that a local function satisfying a series of $(1,m)$-, $(2,m)$-, $(3,m)$-, $\ldots$, $(k,m)$-zeros automatically satisfies the $(1,m)$-, $(1,m+1)$-, $\ldots$, $(1,m+k -1)$-zeros. For $k=n-3$, this would imply a Tr$(\phi^3)$ local ansatz is uniquely fixed by either set of conditions.

\begin{figure}[h]
    \centering
\includegraphics[width=0.6\linewidth]{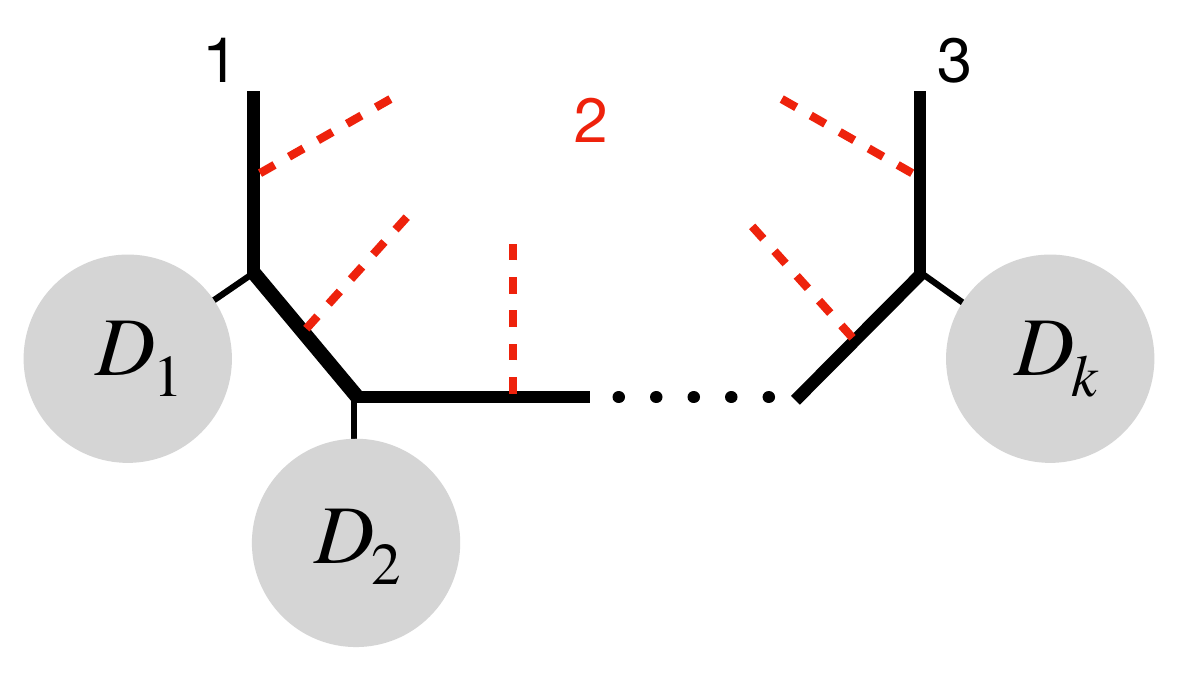}
    \caption{A $D$-subset for leg $2$ is formed by adding leg $2$ in all possible ways to the lower point diagram defined by the sub-diagrams $D_i$. The subset satisfies a $(1,2)$-zero condition if and only if all such diagrams have equal weight.}
    \label{diag}
\end{figure}

To establish this, we use a result from Ref.~\cite{Rodina:2024yfc}, which states that a local function satisfies a $(k,m)$-zero if and only if it can be expressed as a sum over \( D \)-subsets, where each \( D \)-subset independently satisfies the $(k,m)$-zero condition. A \( D \)-subset consists of all \( n \)-point diagrams that can be obtained by attaching legs $(m,m+1,\ldots,m+k-1)$ to an \((n-k)\)-point diagram. This is shown in \cref{diag} for a $(1,2)$-zero.  Moreover, all diagrams within a \( D \)-subset must have equal weight. Therefore, to show that a function satisfies a particular $(1,m)$-zero, we must show that all diagrams of the relevant $D$-subsets have equal weight.

We will demonstrate this for the simplest case, where the $(1,1)$- and $(2,1)$-zeros together imply the $(1,2)$-zero. The proof can easily be generalized to higher cases via induction.

\begin{figure}[h]
    \centering
\includegraphics[width=1\linewidth]{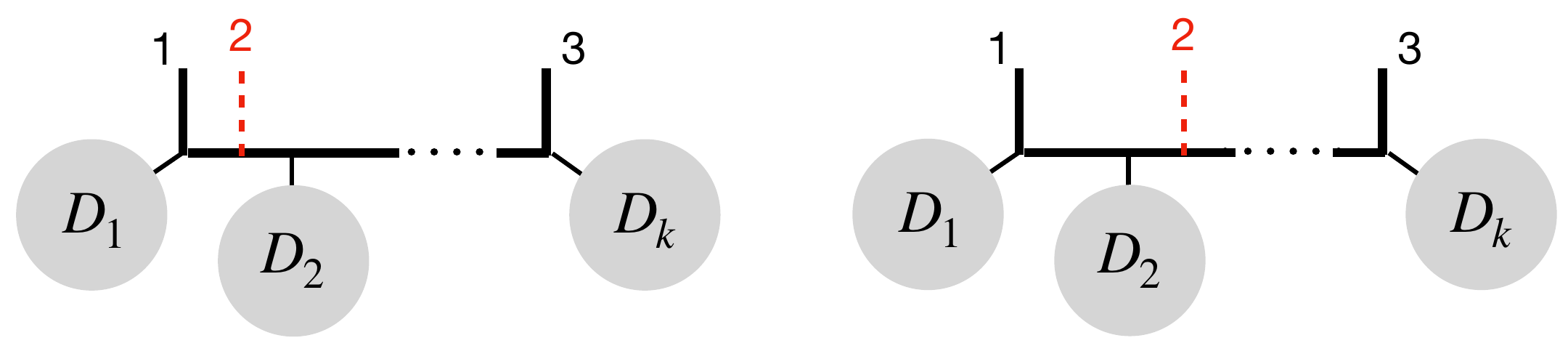}
    \caption{Two diagrams of a particular $D$-subset for leg $2$. We can obtain the $r.h.s.$ diagram from the $l.h.s.$ by removing legs $1$ and $2$, reattaching leg $1$ to its original position, and translating leg $2$. The existence of this $(2,1)$-zero mutation implies the two diagrams have equal weight.}
    \label{fig1A}
\end{figure}

Consider, to start, the diagrams in \cref{fig1A}, which belong to the same $D$-subset for leg $2$ and have legs $1$ and $2$ separated. If this subset satisfies the $(2,1)$-zero, it means all diagrams that can be obtained by removing legs $1$ and $2$ and attaching them anywhere else (respecting ordering and the relative structure of legs $1$ and $2$) must have equal weight. We are therefore free to remove legs $1$ and $2$, place leg $1$ back in its original position, and place leg $2$ in any desired location. This demonstrates that the $(2,1)$-zero enforces equal weights for all diagrams where legs $1$ and $2$ remain separated. 

\begin{figure}[h]
    \centering
\includegraphics[width=1\linewidth]{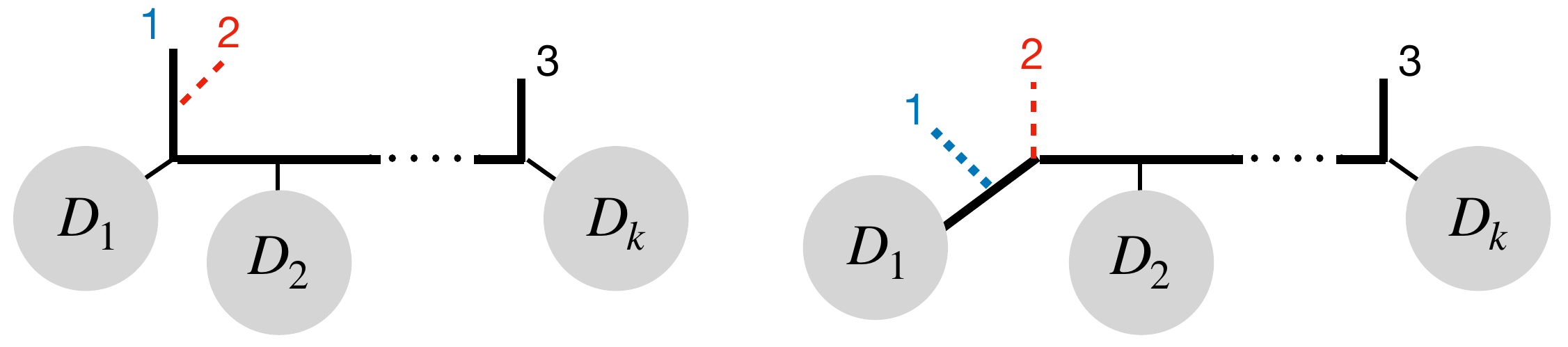}
    \caption{These diagrams are also part of the $D$-subset in \cref{diag}, but are not directly related by a $(2,1)$-zero mutation. They are, however, related by a $(1,1)$-zero mutation.}
    \label{fig2A}
\end{figure}

The complete $D$-subset for leg $2$, however, includes the case where legs $1$ and $2$ form a two-particle pole, as illustrated on the $l.h.s.$ of \cref{fig2A}. Our goal is then to also relate this configuration to the previous diagrams. Since, for this diagram, the $(2,1)$-zero condition only permits the removal and reattachment of the full two-particle pole, we instead now use the (1,1)-zero to move leg $1$ and separate it from leg $2$. This transformation yields the diagram on the $r.h.s.$ in \cref{fig2A}, which now falls under the case just described. So, we've related diagrams with two-particle poles to those with legs $1$ and $2$ separate.

This implies all diagrams of the $D$-subset for leg $2$ in \cref{diag} have equal weight and therefore satisfy the $(1,2)$-zero. The same argument can be applied for all different $D$-subsets, proving that any local function that satisfies a $(1,1)$- and a $(2,1)$-zero automatically satisfies the $(1,2)$-zero. This argument can be generalized by induction, proving our claim.

\section{Appendix D: Fixing $a^+=a^-$}\label{appx.D}
Here, we prove that imposing a final zero on
\eq
\mathcal{M}_{n}=a^+ \ \mathcal{I}_{n}^++a^- \ \mathcal{I}_{n}^-\,,
\eqe
leads to $a^+ = a^-$. To do this, let us impose the $(2, -)$-zero and cut $Y_1^+$. Note that this is exactly the choice where the tree mesh has no overlap with the mountain zero. On this zero, we have $Y_1^+ = -Y_2^-$, and so we get
\eq\label{aa}
    \left(a^+ \mathcal{A}^{(1)}_{n+2} - a^- \mathcal{A}^{(2)}_{n+2}\right)_{(2,-), Y_1^+ = 0} = 0\,,
\eqe
where $\mathcal{A}^{(1)}_{n+2}$ is the tree amplitude resulting from the cut in $Y_1^+$ and $\mathcal{A}^{(2)}_{n+2}$ is that for the cut in $Y_2^-$. Since these $Y$ variables lie at the same point on the loop mesh, the interior of their tree meshes are identical (see \cref{fig:res-kin-mesh}); it is only on the edges that they may differ. 

However, on the support of the zero+cut conditions, it is straightforward to show that the edges also exhibit identical kinematic dependence. As a result, we find that $\mathcal{A}^{(1)}_{n+2}$ and $\mathcal{A}^{(2)}_{n+2}$ are the \textit{same} amplitude on the support of the zero+cut, and thus $a^+ = a^-$ for \cref{aa} to hold.

\section{Appendix E: Unitarity $\Rightarrow$ big mountain zeros}\label{appx.E}

To finish the equivalence from the proof in Sec. 4.1, we need to demonstrate the following statement: if a particular $\mathcal{M}_n = \mathcal{M}^{\star}_n$ does not satisfy some $(i_\star, -)$-zero, then some $a_i \neq a_j$.

We start by imposing the $(i_\star, -)$-zero on $\mathcal{M}^{\star}_n$. Since each term in a local ansatz contains at least one $Y$ variable as a pole, there are two possibilities. If, after imposing the zero, at least one term containing a pole in $Y_i^-$ for $i \neq i_\star$ or $Y_j^+$ for $j \neq i_\star - 1$ in $\mathcal{M}^{\star}_n$ survives, we can freely cut on that variable, giving us the appropriate tree ansatz $\mathcal{B}_{n+2} \neq 0$ on the zero+cut condition. But, as shown Appx. B, the zero+cut condition on $\mathcal{B}_{n+2}$ is equivalent to a tree-level zero. If all the weights in $\mathcal{B}_{n+2}$ were equal, it would necessarily vanish on the this tree-level hidden zero, as demonstrated in Ref.~\cite{Arkani-Hamed:2023swr}. Therefore, the fact that it doesn't means that some of the weights in $\mathcal{M}^{\star}_n$ must be unequal.

In the other case, the only $Y$ variable that survives in $\mathcal{M}^{\star}_n$ on the support of the zero is $Y_{i_{\star}}^- = -Y_{i_{\star} - 1}^+$. Then, cutting $Y_{i_{\star}}^- = -Y_{i_{\star} - 1}^+$ tells us that the \textit{difference} of two tree ans\"atze $\mathcal{B}_{n+2}^{(1)} - \mathcal{B}_{n+2}^{(2)} \neq 0$ on the support of the zero+cut, where $\mathcal{B}_{n+2}^{(1)}$ is what we get from cutting $\mathcal{M}_n^\star$ on $Y_{i_{\star}}^-$ and $\mathcal{B}_{n+2}^{(2)}$ from cutting on $Y_{i_{\star} - 1}^+$, both away from the zero. As was discussed in Appx. D, $\mathcal{B}_{n+2}^{(1)}$ and $\mathcal{B}_{n+2}^{(2)}$ on the support of the zero+cut are the \textit{same} functions of planar variables, just with arbitrarily distinct coefficients $a_i$ and $b_i$. The fact that the difference does not vanish therefore implies that $a_j \neq b_j$ for at least one $j$, and so the weights in $\mathcal{M}_n^\star$ are not all equal. Thus, we have completed the equivalence and shown that a Tr$(\phi^3)$ surface integrand is unitary \textit{if and only if} it satisfies one parity-half of the $2n$ big mountain zeros!

Note that we could have just as well achieved the this result from using the $(i,+)$-zeros. One immediate corollary is then that the local ansatz $\mathcal{M}_n$ satisfies the positive-parity zeros if and only if it satisfies the negative-parity zeros. In this sense, either parity of zeros contains all information on big mountain zeros of the integrand. Another corollary is that the actual unitary one-loop surface integrand $\mathcal{I}_n$ with $a_i = 1$ satisfies the $2n$ big mountain zeros. This is also true for the NLSM surface integrand, which is obtained from Tr$(\phi^3)$ from a $c$-preserving $\delta$-shift. Like in the factorization proof given in Appx. A, we are able to prove the existence of these mountain zeros working entirely in field theory, without using stringy integral techniques.

\section{Appendix F: Integrand examples}\label{appx.E}

In the simplest case of two points, the integrand for Tr$(\phi^3)$ is given by
\eq
\label{eq:two-pt-int}
\mathcal{I}^{\textrm{Tr($\phi^3$)}}_2 = \frac{1}{Y_1^+ Y_2^+} + \frac{1}{Y_1^+ X_{1,1}} + \frac{1}{Y_2^+ X_{2,2}} + (Y^+\leftrightarrow Y^-)\,.
\eqe
One can easily verify by hand that the integrand vanishes on the four big mountains, the $(1,\pm)$- and $(2, \pm)$-zeros.

We also give here the three-point Tr$(\phi^3)$ integrand:
\eqa
\nonumber \mathcal{I}^{\textrm{Tr($\phi^3$)}}_3&=&\left[\frac{1}{Y^+_1 X_{1,1}}\left(\frac{1}{X_{2,1}}+\frac{1}{X_{1,3}}\right)+\textrm{(cyclic)}\right]+\\
\nonumber &&+\left[\frac{1}{Y^+_2 Y^+_3 X_{3,2}}+\textrm{(cyclic)}\right]+\frac{1}{Y^+_1 Y^+_2 Y^+_3}+\\
&&+(Y^+\leftrightarrow Y^-)\,,
\label{eq:three-pt-int}
\eqae
which vanishes, for example, on the $(i,-)$-zeros for $i = 1, 2, 3$ shown in \cref{fig:three-pt-one-loop-mesh}.

More examples (including to higher loop order) can be found in Ref.~\cite{Arkani-Hamed:2024pzc}.

As at tree-level, one may obtain an NLSM integrand through the $\delta$-shift prescription \cite{Arkani-Hamed:2024nhp}. (See Ref.~\cite{Paranjape:2025wjk} for a recent extension.) To do this, we take the Tr$(\phi^3)$ integrand and shift its kinematics $X_{i,j} \to X_{i,j} + \delta_{i,j}$, where $\delta_{e,o} = \delta_{o,e} = 0$ and $\delta_{e,e} = \delta = -\delta_{o,o}$. We also shift $Y_i^\pm$ in the same way, but we first have to choose whether to treat plus as even and minus as odd, or vice versa. After these shifts, we take the limit as $\delta \gg X,Y$ and pick out the leading term, which gives us two possibilities for the NLSM integrand. At two-points, these are
\eq
\label{eq:two-pt-NLSM}
\mathcal{I}^{\textrm{NLSM}}_{2,\pm} =2-\frac{X_{2,2}+Y^\mp_1}{Y^\mp_2}-\frac{X_{1,1}+Y^\pm_2}{Y^\pm_1}\,,
\eqe
where treating plus as even and minus as odd gives us the ``$+$'' configuration. Regardless of this ambiguity, one can verify that both expressions satisfy the same four big mountain zero conditions as \cref{eq:two-pt-int}. This is because both $\delta$-shift conventions preserve the loop-level $c$'s for any multiplicity.

\bibliographystyle{apsrev4-1}
\bibliography{ref2}{}

\end{document}